\documentclass[11pt]{article}

\usepackage{graphicx}
\usepackage{amssymb}

\newcommand{\be}{\begin{equation}}
\newcommand{\ee}{\end{equation}}
\newcommand{\bea}{\begin{eqnarray}}
\newcommand{\eea}{\end{eqnarray}}
\newcommand{\nn}{\nonumber}

\newcommand{\p}{\mbox{$\partial$}}

\begin{document}

\begin{center}

\section*{Mach's principle, Action at a Distance and Cosmology}

\end{center}

\vspace{0.15 in}

\begin{center}
\noindent
{\bf H. Fearn}\\

\vspace{0.1in}
\noindent
\small{Physics Department, California State University Fullerton\\
800 N. State College Blvd., Fullerton CA 92834

\noindent
{\bf email:} hfearn@fullerton.edu}

\vspace{0.2in}
\noindent
{\large \bf Abstract}\\
\end{center}

\noindent
Hoyle and Narlikar (HN) in the 1960's \cite{HN64, HN, HNbook} developed a theory of gravitation which was
completely Machian and used both retarded and advanced waves to communicate gravitational influence between particles. The advanced waves, which travel backward in time, are difficult to visualize and although they are mathematically allowed by relativistic wave equations, they never really caught on. The HN theory reduced to Einstein's theory of gravity in the smooth fluid approximation and a transformation into the rest frame of the fluid.  Unfortunately the theory has been ignored by much of the General Relativity community since it was developed with the static universe in mind. However, it is easy to drop the static universe condition (by dropping the ``C"-field matter creation terms) and then you have a perfectly good theory of gravitation. Hawking \cite{hawking} in 1965 pointed out a possible flaw in the theory. This involved integrating out into the distant future to account for all the advanced waves which might influence the mass of a particle here and now. Hawking used infinity as his upper time limit and showed the integral was divergent. We would like to point out that since the universe is known to be expanding, and accelerating, the upper limit in the advanced wave time integral should not be infinite but is bounded by the Cosmic Event Horizon. This event horizon $H_e$ represents a barrier between future events that can be observed and those which cannot. We maintain that the advanced integral is in fact finite when the cosmic event horizon is taken into account, since the upper limit of the advanced wave integral becomes $H_e/c$. Hawking's objection is no longer valid and the HN theory becomes a working theory once again.

\vspace{0.2in}
\noindent
\subsubsection*{History of Mach's principle}

Mach's principle was the name Einstein gave to the proposition that the origin of inertia and inertial forces is the gravitational interaction in 1918.  The idea that this might be the case had occurred to him at least nearly four years earlier, for he elaborated it in a short paper where he considered the gravitational interaction of a spherical shell of material and a point mass located at the center of the shell, \cite{shell}.  What he found was that the mass shell increases the inertial mass of the particle inside the shell. According to Einstein, this suggested that the \emph{entire} inertial mass of the material particle is an effect of the presence of all other masses in the universe, based on a gravitational interaction with the latter. Einstein showed that with the particle stationary and the shell accelerated the particle would be dragged along by the shell. If the mass shell was of mass $M$ and radius $R$ and the particle of small mass $m$, if the particle was at rest and the shell accelerated by $a_M$ then the resulting force on the particle would be,
\be
f = -\frac{3}{2} \left( \frac{ GMm}{Rc^2} \right) a_M \;\; .
\ee
According to Einstein, the force $f$ must be exerted on the particle for it to remain at rest. Thus $-f$ is the force exerted (induced by gravitation) on the particle by the shell which has an acceleration $a_M$. The particle is therefore dragged along by the shell. We may also consider the shell to be at rest and the particle to be accelerating at $a_m$ with some external force $f$. In this case the force on the shell $F$ is,
\bea
F &=&  -\frac{3}{2} \left( \frac{ GMm}{Rc^2} \right) a_m \nn \\
f &=& \left( m +  \frac{ GMm}{Rc^2} \right) a_m
\eea
where the term in brackets in the last equation is the inertial mass of the particle. If we take $F+f=0$ so that energy-momentum is conserved we get,
\bea
\left( m - \frac{GMm}{2 R c^2} \right) a_m &=& 0 \nn \\
\Rightarrow  \frac{GM}{2R} &=& c^2 \;\; .
\eea

\noindent
Although Einstein was considering a weak field limit, Lynden-Bell later showed a similar result for dragging of a central test particle by a massive accelerating shell in the case of strong gravitational fields, \cite{bell}.\\

\noindent
Einstein understood in 1912 that his speculations on gravity and inertia were just that, speculations, as he did not then possess a ``serviceable dynamical theory of the gravitational field''.  Several years later, with his creation of general relativity theory, he did have that theory of the dynamical gravitational field.  And not long thereafter, he was talking about the ``relativity (and origin) of inertia'' and ``Mach’s principle''.  His most extensive comments on Mach's ideas on inertia, amplified by him, appear in The Meaning of Relativity \cite{albertbk}, the published version of lectures he gave at Princeton in the spring of 1921.  He shows that in the linear, weak field approximation, general relativity does in fact predict the sorts of effects one might expect if the origin of inertia and inertial forces are gravitational.  But he makes no claim that inertia is exclusively a gravitational phenomenon – though it is easy to read between the lines that this was his belief, for he asserts that the universe must be closed and the source of inertia the material contents of the universe.  The reason for Einstein’s reticence seems to have been the cosmological information available at the time, which would make the scalar part of the total gravitational potential, GM/R, a millionth or so of the value of the square of the vacuum speed of light.  A minute fraction of the value needed for inertia to be attributable to gravity alone.\\

\noindent
Also, Einstein’s views on inertia had been tempered by an exchange with Willem deSitter, who had shown convincingly that the field equations of general relativity admit solutions that are manifestly incompatible with any reasonable definition of Mach’s principle, namely an empty universe.  Moreover, Einstein had come to appreciate that, as inertial forces are acceleration dependent forces, they had at least one of the signatures of radiative interactions.  This had led him to draw a distinction between, as he called it, ``the relativity of inertia'' and ``Mach's principle''.  Mach's principle, which he took to encompass the radiative nature of the presumed interaction between test particles and the rest of the ``matter'' in the universe, he claimed was based on ``action at a distance'' as inertial forces are experienced instantly on the application of ``external'' forces.  In contradistinction, the relativity of inertia merely required that inertia and inertial forces be dependent on the presence of a field to act on accelerating objects, and thus could be encompassed by his theory of gravity, general relativity.\\

\noindent
The type of action at distance that Einstein was talking about was the Newtonian type: instantaneous communication of effects over finite distances.  The modern concept of action at a distance – one that is consistent with the principle of relativity – is different.\\

\noindent
The modern version of action at a distance was first introduced in the 1920s by Hugo
Tetrode \cite{tetrode} and Adriaan Fokker \cite{fokker}.  It was Fokker's action that Dirac invoked when he wrote his classic paper on the self-energy and radiation reaction issues for classical electrons in 1938 \cite{dirac}.
The distinctive feature of this work is that advanced, as well as retarded solutions of the equations of electrodynamics are admitted.  This makes seemingly instantaneous action possible as advanced waves can be used to communicate the influence of future events to the present.
The idea is as follows, consider a single electron undergoing acceleration. The field surrounding the electron can be thought of in two parts, the outgoing and incoming. The actual field surrounding the electron is the usual retarded Lienard Wiechert potentials and any incident field on the electron.
\be
F^{\mu \nu}_{\mbox{act}} = F^{\mu \nu}_{\mbox{ret}} + F^{\mu \nu}_{\mbox{in}}
\ee
\noindent
The Maxwell 4-potential wave equation allows for advanced solutions, which are the same form as retarded only they go backward in time. We could equally well describe the actual field surrounding the electron by
\be
F^{\mu \nu}_{\mbox{act}} = F^{\mu \nu}_{\mbox{adv}} + F^{\mu \nu}_{\mbox{out}}
\ee
where the $F^{\mu \nu}_{\mbox{out}}$ is the total field leaving the electron. The difference
between the outgoing waves and the incoming waves is the radiation produced by the electron due to
its acceleration.
\be
F^{\mu \nu}_{\mbox{rad}} = F^{\mu \nu}_{\mbox{out}} - F^{\mu \nu}_{\mbox{in}}=
F^{\mu \nu}_{\mbox{ret}} - F^{\mu \nu}_{\mbox{adv}}
\ee
\noindent
In the appendix of Dirac's paper, it is shown that this equation gives exactly the well known relativistic result
for radiation reaction which can be found in standard text books on electromagnetism. John Wheeler and Richard Feynman \cite{WF} took it up and worked through how it would have to be formulated to correspond to observation.  They called it ``absorber'' electrodynamics, for the condition that makes the theory work is complete absorption of all retarded waves propagating into the future – and the generation of advanced waves as the absorber is acted upon that combine with the retarded waves to produce what appears to be a purely retarded interaction. Wheeler-Feynman resort to a suggestion made by  Tetrode \cite{tetrode} and later by Lewis \cite{lewis} which was to abandon the concept of electromagnetic radiation as a self interaction and instead interpret it as a consequence of an interaction between the source accelerating charge and a distant absorber. The absorber idea has the four following basic assumptions, \cite{WF},

\begin{quote}
\noindent\emph{
(1) An accelerated point charge in otherwise charge-free space does not radiate electromagnetic energy.\\
(2) The fields which act on a given particle arise only from other particles.\\
(3) These fields are represented by 1/2 the retarded plus 1/2 the advanced Lienard-Wiechert solutions of Maxwell's equations. This force is symmetric with respect to past and future.\\
(4) Sufficiently many particles are present to absorb completely the radiation given off by the source.}\\
\end{quote}

\noindent
Now Wheeler-Feynman considered an accelerated charge located within the absorbing medium. A \emph{disturbance}  travels outward from the source. The absorber particles react to this disturbance and themselves generate a field half advanced and half retarded. The sum of the advanced and retarded effects of all the charged particles of the absorber, evaluated near the source charge give an electromagnetic field with the following properties, \cite{WF};\\

\begin{quote}

\emph{
\noindent
(1) It is independent of the properties of the absorbing medium.\\
(2) It is completely determined by the motion of the source.\\
(3) It exerts on the source a force which is finite, is simultaneous with the moment of acceleration, and is just sufficient in magnitude and direction to take away from the source the energy which later shows up in the surrounding particles.\\
(4) It is equal in magnitude to 1/2 the retarded field minus 1/2 the advanced field generated by the accelerated charge. In other words, the absorber is the physical origin of Dirac's radiation field...\\
(5) This field combines with the 1/2 retarded, 1/2 advanced field of the source to give for the total disturbance the full retarded field which accords with experience.}\\
\end{quote}

\noindent
It turns out that absorber theory could not account for the electron self energy, electrons do in fact interact with themselves. For this reason Wheeler and Feynman both gave up on it. Relativistic wave equations do have both retarded and advanced solutions and the mere fact that the infinite electron self energy could not be accounted for should not be the sole reason to dismiss absorber theory. Retarded and advanced signals are shown in figure 1 using a Minkowski diagram to see the forward and backward travel in time.

\begin{figure}
  \begin{center}
   \includegraphics[height=2in, width=1.75in]{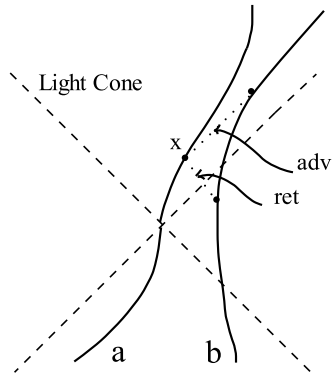}\\
   \caption{Particle {\bf b} interacts with particle {\bf a} at point x via retarded and advanced waves. The mass field
   (gravitational waves) travel at speed c dotted lines. The light cone is shown dashed and the world lines of both particles are solid lines.}\label{fig1}
  \end{center}
\end{figure}

\noindent
However, absorber theory never caught on, for it is highly counter-intuitive in several ways. Indeed, Feynman recounts that when he presented the theory to a Princeton Physics Department colloquium \cite{drum}:

\begin{quote}
 Wolfgang Pauli who was sitting next to Einstein, said: `I do not think this theory can be right because of this, that and the other thing.' . . . At the end of this criticism, Pauli said to Einstein, `Don't you agree, Professor Einstein?  I don't believe this is right, don't you agree, Professor Einstein?'  Einstein said, `No,' in a soft German voice that sounded very pleasant to me, very polite.  `I find only that it would be very difficult to make a corresponding theory [i.e., an action at a distance theory] for gravitational interactions.'
\end{quote}

\noindent
 Einstein didn't need to try to construct such a theory, for he was certain that inertia was already accounted for as a gravitational phenomenon in general relativity. But others weren't so convinced.

\noindent
The golden age of arguments about Mach's principle followed, with recurrences at intervals of roughly 20 years since.  Given what we now know about cosmology, it is clear that Einstein’s belief that the relativity of inertia was built into general relativity was justified.  Observed spatial flatness at cosmic scale ensures that $GM/R \approx c^2$, and as Einstein discovered in 1912, that's what's required.  But general relativity does not encompass the version of Mach's principle that includes inertial actions as radiative in nature.\\

\noindent
Fred Hoyle and Jayant Narlikar \cite{HN64,HN,HNbook} tried to generalize general relativity to correct this perceived defect in the mid-1960s by constructing an action-at-a-distance theory of gravity along the lines of the Wheeler-Feynman absorber theory of electrodynamics.  The theory did not find much favor at the time, no doubt in part due to Hoyle's ardent support for steady state cosmology.  But there was a more fundamental problem, quickly spotted by the then graduate student Stephen Hawking \cite{hawking}.  The analogy between electrodynamics and gravity is not exact.  Electromagnetic fields can be ``screened'' by matter.  So, they will be absorbed if there is sufficient matter along the future light cone, as is evidently the case, see below. This is not true for the gravitational interaction as it is normally understood.  While energy can be extracted (absorbed) from gravitational waves, the underlying gravitational field is NOT absorbed.  That is, gravity cannot be ``screened'' by any known substance or process.  Not yet anyway.\\

\noindent
In the early 1970's Partridge \cite{partridge} attempted to detect advanced waves indirectly by measuring the output intensity of a microwave transmitter. He aimed the transmitter at a perfect absorber in the laboratory and noted the intensity output, then he aimed the transmitter at the zenith (into space). The idea was that the universe was not such a good absorber of microwaves and so there should be a drop in the intensity of the microwave transmitter. According to Wheeler Feynman absorber theory, if there is not perfect absorption then an excited atom would not emit. Unfortunately, Partridge failed to take into account the expansion of the universe and the Doppler red shift which would eventually ``stretch" his microwaves to $21$cm length and then they would inevitably be absorbed by interstellar hydrogen. He got inconclusive results. Later in the 80's John Cramer \cite{cramer1} tried using neutrinos, from beta decay, in a similar experiment. The idea again was to detect advanced waves by measuring an intensity change when the neutrino beam was scanned across the sky. Cramer's hypothesis was that the neutrinos would be selectively absorbed in the directions of black holes (BH's). The intensity of neutrinos (and electrons) from his radioactive beta source should have been greater when the direction corresponded to the direction of a BH. This experiment was also inconclusive.\\

\noindent
Hawking \cite{hawking} was quick to see that this meant that all gravitational influences propagating at speed c do so into the infinitely far/distant future – notwithstanding the expansion of the universe in cosmological models of the time (1965)– and if the retarded influences can get there (where-ever), the advanced influences can get back to the present irrespective of ``horizons'' of most sorts that might intervene. This means that the integral of all of the advanced contributions from those propagating influences will diverge \cite{jim}. Unless one’s theory of gravity allows for an ``absorption'' or ``screening'' mechanism, it would seem that this objection to action at a distance gravity is insuperable. The mathematical details will be shown later.\\

\noindent
But such a view would be too pessimistic.  What is needed is a way to cut the gravitational interaction off at a finite distance/time.  Note that closing the universe, as Einstein would do, doesn't solve this problem, for the influence can simply recirculate in a closed universe ad nauseam.  But accelerating expansion can do the trick.\\

\subsubsection*{Hoyle Narlikar Theory}

\noindent
The direct-particle action principle from which Hoyle and Narlikar (HN) derive their theory of gravitation, yields Einstein's general gravitational theory in the limit of a smooth fluid. It is also necessary to transform to the rest frame of the fluid.
The original HN theory has built into it a matter creation field called the ``C-field" which allows for a static universe. The creation field adds in matter as the universe expands so that the matter density remains constant. We suggest that the matter creation field be dropped and do not consider it here. The basic assumptions of the theory are as follows;\\

\noindent
\emph{(1) The mass $m(x_a)$ (mass at position $x_a$) must become a direct particle field, it must arise from all the other mass in the universe.\\
(2) Since mass is scalar we expect it to arise through a scalar Greens function.\\
(3) The action must be symmetric between any pairs of particles, \cite{HNbook}.\\ }

Let each particle `b' give rise to a mass-field. Denote this field at a general point $x$ by $m^{\mbox{\tiny(b)}}(x)$. At any point $x_a$ on the path of particle-a, we have
$m^{\mbox{\tiny(b)}}(x_a)$ as the contribution of particle-b to the mass of particle-a at the position $x_a$. Summing for all b particles ,
\be
m(x_a) =  \sum_{b} m^{\mbox{\tiny (b)}} (x_a) = \sum_b \int G^* (x_a,x_b) dx_b    \label{massa}
\ee
this gives the mass at point $x_a$ due to all particles including those at position $x_a$. The full mass field is the sum of
half retarded and half advanced fields,

\noindent
where $G^*(x_a,x_b) = (1/2)( G(x_a,x_b) +  G(x_b,x_a) )$ is a time-symmetric Green's function.
\be
m(x) = \frac{1}{2} \left( m_{\mbox{\tiny ret}} + m_{\mbox{\tiny adv}} \right)
\ee
The m-field satisfies the following wave equation,
\be
\Box m + \frac{1}{6} R m = N
\ee
where $N>0$ is the density of world lines of particles.
The derivation of the new field equations is fairly long and can be found in detail in the book by Hoyle and Narlikar \cite{HNbook} and summarized by Hawking \cite{hawking}. The action $I$ used can be written in terms of $m$ or in terms of the $G^*(x_a,x_b)$ as follows;
\bea
I &=& \sum_{a\neq b}\sum \int \int G^*(x_a,x_b) dx_a dx_b \nn \\
&=& - \int m ds
\eea
where the mass $m$ is thought of as being due to the interaction of one particle with all the other mass energy in the universe. The field equations are derived most readily by using the $G^*$ equation and the equations of motion are most easily derived using the variable mass $m$. Here, we give a new short derivation of the equation of motion,
in slightly modified notation which we hope will be easy to follow.\\


\noindent
{\bf Equations of Motion for HN-theory}\\

\noindent
Starting from the line element,
\bea
ds^2 &=& g_{\mu\nu} dx^\mu dx^\nu \nn \\
2 ds \delta (ds) &=& \delta g_{\mu\nu} dx^\mu dx^\nu + 2 g_{\mu\nu} dx^\mu \delta ( dx^\nu ) \nn \\
\delta ( ds ) &=& \left[ \frac{1}{2} \delta g_{\mu \nu } \dot{x}^\mu \dot{x}^\nu  + g_{\mu \nu} \dot{x}^\mu \frac{d}{ds} (\delta x^\nu ) \right] ds
\eea

\noindent
Now the action for mass $m$ at position $x$ can be simply written as,
\bea
I &=& - \int m ds \nn \\
\delta I &=& - \int \left[ \delta ( m ) ds + m \delta (ds) \right] \nn \\
&=& -\int \left[ \frac{\p m}{\p x^\lambda }\delta x^\lambda + \frac{m}{2} \delta g_{\mu \nu} \dot{x}^\mu \dot{x}^\nu +
m g_{\mu \nu} \dot{x}^\mu \frac{d}{ds} ( \delta x^\nu ) \right] ds
\eea

\noindent
Integrate the last term by parts and switch dummy variable $\nu \rightarrow \lambda$ we get,

\bea
\delta I &=& - \int \left[\frac{\p m}{\p x^\lambda }\delta x^\lambda +
\frac{m}{2} \frac{\p g_{\mu \nu}}{\p x^\lambda }  \dot{x}^\mu \dot{x}^\nu \delta x^\lambda -
\frac{d}{ds} ( m g_{\mu\lambda} \dot{x}^\mu ) \delta x^\lambda \right] ds \nn \\
&=& - \int \left[ \frac{\p m}{\p x^\lambda } + \frac{m}{2} \frac{\p g_{\mu \nu}}{\p x^\lambda }  \dot{x}^\mu \dot{x}^\nu -
\frac{d}{ds} ( m g_{\mu\lambda} \dot{x}^\mu ) \right] \delta x^\lambda ds =0
\eea

\noindent
For this integral to be zero for any arbitrary $\delta x^\lambda$ then the term in the square brackets must be zero, hence

\bea
\frac{d}{ds} ( m g_{\mu\lambda} \dot{x}^\mu ) &=& \frac{m}{2} \frac{\p g_{\mu \nu}}{\p x^\lambda }  \dot{x}^\mu \dot{x}^\nu
+ \frac{\p m}{\p x^\lambda } \nn \\
\frac{dm}{ds} g_{\mu \lambda} \dot{x}^\mu + m \left( g_{\mu \lambda} \frac{d \dot{x}^\mu }{d s} + g_{\mu\lambda, \nu} \dot{x}^\mu \dot{x}^\nu \right) &=& \frac{m}{2} g_{\mu \nu, \lambda}  \dot{x}^\mu \dot{x}^\nu
+ \frac{\p m}{\p x^\lambda }
\eea

\noindent
where we may make the $g_{\mu\lambda, \nu}$ term symmetric in $\mu,\nu$ as follows,
\be
g_{\mu \lambda} \frac{d}{ds} ( m \dot{x}^\mu ) = \frac{m}{2} ( g_{\mu \nu, \lambda} - g_{\mu\lambda , \nu} -
g_{\nu \lambda, \mu} ) \dot{x}^\mu \dot{x}^\nu + \frac{\p m}{\p x^\lambda}
\ee
Then using the definition for the Christoffel symbol $\Gamma_{\lambda \mu, \nu}$ and multiplying throughout by
$g^{\sigma \lambda}$
we get,

\bea
\frac{d}{ds} ( m \dot{x}^\sigma )  + m ( g^{\sigma \lambda } \Gamma_{\lambda \mu, \nu} ) \dot{x}^\mu \dot{x}^\nu - g^{\sigma \lambda} \frac{\p m}{\p x^\lambda} &=& 0 \nn \\
\frac{d}{ds} ( m \dot{x}^\sigma )  + m \Gamma_{\mu \nu}^\sigma \dot{x}^\mu \dot{x}^\nu -
g^{\sigma \lambda} \frac{\p m}{\p x^\lambda} &=& 0
\eea

\noindent
Written for mass $m_a$ at position $x_a$ the equation of motion becomes,
\be
\frac{d}{d\tau} \left( m_a \frac{dx_a^\mu}{d\tau}  \right) + m_a \Gamma_{\nu\lambda}^\mu \frac{dx_a^\nu}{d\tau}\frac{dx_a^\lambda}{d\tau} - g^{\mu\nu} \frac{\p m_a}{\p x_a^\nu} = e_a \sum_{b\neq a} F^{\mbox{\tiny (b)}\mu}_\nu \frac{dx_a^\nu}{d\tau}
\ee
where the Lorentz force has been included on the right for completeness.The world-lines of particles are not in general geodesics in the new theory.\\

\noindent
The familiar energy momentum tensor is easily derived, see Hoyle and Narlikar's book \cite{HNbook} p112. This follows from using  $d\tau^2 = g_{\mu\nu} dx^\mu dx^\nu$ and thus
$\delta (d\tau) = \delta g_{\mu\nu} \dot{x}^\mu \dot{x}^\nu d\tau$ which leads to,

\bea
-\sum_a \int m(x_a) \delta ( d \tau ) &=&  -\sum_a \int m(x) \delta g_{\mu\nu} \dot{x}^\mu \dot{x}^\nu
\delta^4( x-x_a) d\tau \nn \\
&=&-\int_V T^{\mu\nu} \delta g_{\mu\nu} [-g]^{1/2} d^4x = +\int_V T_{\mu\nu} \delta g^{\mu\nu} [-g]^{1/2} d^4x \nn \\
\mbox{  where  } T^{\mu\nu} &=& \sum_a  \delta^4 (x - x_a) [ -g]^{-1/2} m(x) \dot{x}^\mu \dot{x}^\nu d\tau \;\; .
\eea

\noindent
The field equations can be derived from a variation of the scalar Green's function. A brief summary is given below, it is a lengthy calculation and some steps are missing. A slightly modified notation has been used which is hopefully more familiar,
\bea
\delta I &=& - \frac{1}{2} \delta \left[ \sum_a \int m(x_a) d\tau \right] \nn \\
&=& -\frac{1}{2} \sum_a m(x_a) \delta (d\tau) - \frac{1}{2} \sum_a \int \delta m(x_a) d\tau \nn \\
&=& - \frac{1}{2} \int_V  T^{\mu\nu} \delta g_{\mu\nu} [-g]^{1/2} d^4x -
\frac{1}{2}\sum_a \sum_b \int \int \delta G(x_a,x_b) d\tau d\tau' \nn \\
&=& +\frac{1}{2} \int_V T_{\mu\nu} \delta g^{\mu\nu} [-g]^{1/2} d^4x
- \frac{1}{2}  \int_V \delta ([-g]^{1/2} g^{\mu\nu})
\left[  \frac{\p m^{\mbox{\tiny adv}}}{\p x^\mu }\frac{\p m^{\mbox{\tiny ret}}}{\p x^\nu} \right] d^4x \nn \\
& & + \frac{1}{12}  \int_V \delta ( R [-g]^{1/2} ) \left[ m^{\mbox{\tiny adv}} m^{\mbox{\tiny ret}} \right] d^4x
\eea
where we have used the following useful identity \cite{HNbook} p 113,
\be
\delta [-g]^{1/2} = -\frac{1}{2} g_{\mu\nu}\delta g^{\mu\nu} [-g]^{1/2} \label{del} \;\; .
\ee
\noindent
The $\delta ( R [-g]^{1/2} )$ term in $\delta I$ can be expanded also as follows;
\bea
\frac{1}{6} \delta ( R [-g]^{1/2} )  &=& \frac{1}{6} \delta (R_{\mu\nu} g^{\mu\nu} [-g]^{1/2} ) \nn \\
&=& \frac{1}{6} \left[ \delta ( g^{\mu\nu} R_{\mu\nu} ) [-g]^{1/2} + R \delta( [-g]^{1/2} ) \right]
\eea
The variation of the action $\delta I$ then becomes,
\bea
\delta I &=& +\frac{1}{2} \int_V T_{\mu\nu} \delta g^{\mu\nu} [-g]^{1/2} d^4x \nn \\
& & +\frac{1}{2} \left[ \left( \frac{1}{2} g_{\mu\nu} \delta g^{\mu\nu} [-g]^{1/2}\right) g^{\alpha\beta}
\frac{\p m^{\mbox{\tiny adv}}}{\p x^\alpha }\frac{\p m^{\mbox{\tiny ret}}}{\p x^\beta} \right] \nn \\
& & \;\;\;\;\;\;\;\;\;\; - [-g]^{1/2} \delta g^{\mu\nu} \frac{1}{2}\left[
\frac{\p m^{\mbox{\tiny adv}}}{\p x^\mu }\frac{\p m^{\mbox{\tiny ret}}}{\p x^\nu} +
\frac{\p m^{\mbox{\tiny adv}}}{\p x^\nu }\frac{\p m^{\mbox{\tiny ret}}}{\p x^\mu} \right] \nn \\
& & +\frac{1}{6} \int_V  \left( R_{\mu\nu} - \frac{1}{2} R g_{\mu\nu} \right)
\delta g^{\mu\nu} [-g]^{1/2}\left[ m^{\mbox{\tiny adv}} m^{\mbox{\tiny ret}} \right] \nn \\
& & +\frac{1}{2} \int \theta_{\mu\nu} \delta g^{\mu\nu} [-g]^{1/2} d^4x = 0 \; .
\eea

\noindent
The field equations are then seen to be \cite{HNbook},
\bea
T_{\mu\nu} &+& \theta_{\mu\nu} + \frac{1}{6} ( R_{\mu\nu} - \frac{1}{2} g_{\mu\nu} R )m^{\mbox{\tiny adv}} m^{\mbox{\tiny ret}} \nn \\
&-&\frac{1}{2} \left[
\frac{\p m^{\mbox{\tiny adv}}}{\p x^\mu }\frac{\p m^{\mbox{\tiny ret}}}{\p x^\nu} +
\frac{\p m^{\mbox{\tiny adv}}}{\p x^\nu }\frac{\p m^{\mbox{\tiny ret}}}{\p x^\mu}  -
g_{\mu\nu} g^{\alpha\beta}\frac{\p m^{\mbox{\tiny adv}}}{\p x^\alpha }\frac{\p m^{\mbox{\tiny ret}}}{\p x^\beta}
\right] = 0 \nn \\
\eea
what remains is to expand $\theta_{\mu\nu}$ in its full glory. After a few pages of algebra we get,
\noindent
\be
\theta_{\mu\nu} = - \frac{1}{6} \left[ g_{\mu\nu} \Box^2 ( m^{\mbox{\tiny adv}} m^{\mbox{\tiny ret}} ) -
( m^{\mbox{\tiny adv}} m^{\mbox{\tiny ret}} )_{,\mu\nu} \right]
\ee
where $\Box^2$ is the wave equation $\p_\mu \p^\mu $. These equations will reduce to the usual Einstein field equations in the limit of a smooth fluid. \cite{HNbook}.\\

\vspace{0.2in}

\noindent
{\bf Details of Hawking's Argument for a divergent advanced mass}\\

\noindent
In 1965 Hawking \cite{hawking} claimed that the HN theory was incompatible with an expanding universe since the advanced mass field would be infinite.  He suggested that a possible way out of this problem was to include negative mass particles in the derivation. At the time it was not known that the universe was accelerating in its expansion. The book by Hoyle and Narlikar does have the negative mass terms worked out but they are not needed to show that the advanced component of the mass field does not diverge. Let us expand on the problem here;\\

\noindent
The Robertson Walker (RW) line element has the standard form,
\be
ds^2 = dt^2 - R^2 (t) \left[ \frac{ dr^2}{1 - kr^2} + r^2 ( d\theta^2 + \sin^2 \theta d\phi^2 ) \right] \;\; ,
\ee
written in comoving coordinates where $c=1$. The coordinates of a massive body are fixed in this system the only thing changing is the scale factor $R(t)$. One can imagine two masses stuck onto a balloon, and then the inflation of the balloon represents the expansion of the universe, \cite{ajp}. The geometric factor $k$ is +1 for a closed universe , 0 for flat and -1 for an open universe. The RW metric is applied to the Einstein field equations to obtain the Friedmann equations \cite{ajp,weinberg}. The space-space term and the time-time (acceleration) term become respectively,
\bea
R \ddot{R} + 2 \dot{R}^2 + 2k &=& 4 \pi G (\rho - p ) R^2 \nn \\
3 \ddot{R} &=& - 4 \pi G ( \rho + 3 p) R
\eea
Eliminating $\ddot{R}$ between these two equations gives,
\bea
H^2 \equiv \left(\frac{\dot{R} }{R} \right)^2 = \frac{ 8 \pi G}{3 } \rho - \frac{k }{R^2} \;\; .\label{einstein}
\eea
For Einstein-de Sitter (flat) space we take $k=0$.
The RW line element is easily seen to be conformal to Minkowski space by defining a new time coordinate, \cite{hawking} we have
\bea
ds^2 &=& \Omega^2 [ d\tau^2 - dr^2 + r^2 d\theta^2 + r^2 \sin^2 \theta d\phi^2 ] \nn \\
&=& \Omega^2 \eta_{\mu\nu} dx^\mu dx^\nu
\eea
where $\eta_{\mu\nu}$ is the flat space metric.
The conformal function is given by $\Omega(t) = R(t)$ so that mass in the Minkowski frame is $m^* = R m$ .
If we write $n$ for the particle density in the original RW frame then the corresponding density $n^*$ in the Minkowski frame is $n^* = R^3 n=L^{-3}$.  Using the Eq. (\ref{einstein}) and writing $\rho= n m$ and using $8 \pi G = 6/m^2$ from HN-theory \cite{HNbook}, we have,
\bea
\frac{\dot{R}^2}{R^2} &=& \frac{2 \rho}{  m^2 } = \frac{ 2 n }{m} \nn \\
\dot{R}^2 &=& \frac{2 n R^3 }{m R} = \frac{2 n^*}{mR} = \left( \frac{2 L^{-3}}{m } \right) \frac{1}{R}
\eea
the term in brackets is a constant. Integrating and using $R=0$ at $t=0$ we find
\bea
\int_0^t R^{1/2} dR &=& \int_0^t \left( \frac{2 L^{-3}}{m } \right)^{1/2} dt \nn \\
R(t) &=& \left( \frac{2 L^{-3}}{m } \right)^{1/3} \left( \frac{3}{2} \right)^{2/3} \left( t \right)^{2/3}
\eea
Now we can write $R$ in terms of $\tau$ by using the integral
\bea
\tau &=& \int_0^t \frac{dt}{R(t)} = 3 \left( \frac{2}{3} \right)^{2/3} \left( \frac{m}{2 L^{-3}} \right)^{1/3} t^{1/3} \nn \\
\frac{1}{2} \tau &=& \left( \frac{m}{2 L^{-3}} \right)^{1/3} \left( \frac{ 3 t}{2} \right)^{1/3}
\eea
we find $R$ in terms of $\tau$ to be
\be
R(\tau) = \frac{1}{4} \tau^2 \left( \frac{ 2 L^{-3}}{m} \right)
\ee
and hence define the mass in the Minkowski frame to be
\be
m^* = m R(t)= \frac{1}{2} \tau^2  L^{-3} \;\; .
\ee
\noindent
Thus we can define,
\bea
R(\tau) &=& \frac{1}{4} \tau^2 \left( \frac{2 L^{-3}}{m} \right)=\frac{\tau^2}{T^2} \nn \\
T^2 &=& \left( \frac{2 m}{ L^{-3}} \right)
\eea

\noindent
Substituting the RW metric into the energy conservation equation $T^{\mu\nu}_{;\nu}=0$, the time component gives,
\be
\dot{p} R^3 = \frac{d}{dt} \left( R^3 [ \rho + p] \right)
\ee
and this last equation can be written as \cite{weinberg},
\be
\frac{d}{dt} ( \rho R^3 ) = -3 p R^2 \;\; .
\ee

From the last equation it is clear that if the energy density of the universe is dominated by matter with negligible pressure then for $p << \rho$ we find $ \rho_{\mbox{\tiny mat}} \propto R^{-3}$. (If the universe is dominated by relativistic particles like photons then $ p =1/3 $ which gives $ \rho_{\mbox{\tiny rad}} \propto R^{-4} $.)
Hawking points out that for a matter dominated universe without creation, eg. the Einstein-de Sitter universe, the density of world lines of particles is $N = n R^{-3}$ where $n$ is a constant. The scale factor $R(t) \propto (t)^{2/3}$ as shown above and $\Omega = R(\tau) = ( \tau/ T)^2$ \cite{hawking}.\\

\noindent
From HN-theory \cite{HNbook} the Greens function obeys the wave equation,
\be
\Box G^* (x_a,x_b) + \frac{1}{6} R G^*(x_a,x_b) = \frac{\delta^4 (x_a,x_b)}{ \sqrt{-g}} \;\; .
\ee
\noindent
This is a flat space Green function of the type \cite{hawking}
\be
G^*(\tau_1, 0; \tau_2 ,r) = \frac{ 1}{ 8 \pi \Omega(\tau_1)\Omega(\tau_2)} \left[
\frac{\delta ( r - \tau_2 + \tau_1 )}{ r } + \frac{\delta ( r + \tau_2 - \tau_1 )}{ r } \right]
\ee
\noindent
The mass field or `m'-field is given by
\be
m(\tau_1) = \int G^* N \sqrt{ -g} dx^4 = \frac{1}{2} \left( m_{\mbox{\tiny ret}} + m_{\mbox{\tiny adv}} \right)
\ee
\noindent
where $ dx^4 = 4 \pi \Omega^4(\tau_2) r^2 dr d\tau_2 $.

For the retarded mass field we may write,
\be
m_{\mbox{\tiny ret }}(\tau_1) = \frac{1}{\Omega(\tau_1)} \int \frac{ N \delta ( r + \tau_2 - \tau_1 )
\Omega^3(\tau_2 ) }{ 4 \pi r } 4 \pi r^2 dr d\tau_2
\ee
where the $R$ terms from $N = nR^{-3}$ and $\Omega^3$ cancel, and the $r$ integral can be performed  using the delta function to give,
\be
 m_{\mbox{\tiny ret }}(\tau_1) = \frac{T^2}{\tau_1^2} \int_0^{\tau_1}  n (\tau_1 - \tau_2) d\tau_2
  = \frac{1}{2} n T^2 = \frac{n}{L^{-3}} m
\ee
where $\tau_1 $ is the current age of the universe and the integral is over the past light cone. We have used our previous result for $T^2$. Also $R(\tau) = \Omega(\tau)$.\\

\noindent
For the advanced waves only we may write,
\be
m_{\mbox{\tiny adv }}(\tau_1) = \frac{1}{\Omega(\tau_1)} \int
\frac{ N \delta( r -\tau_2 + \tau_1) \Omega^3(\tau_2 ) }{ 4 \pi r } 4 \pi r^2 dr d\tau_2
\ee
\noindent
which integrated over the future light cone we find,
\be
m_{\mbox{\tiny adv}} (\tau_1) = \left( \frac{T}{\tau_1 } \right)^2 \int_{\tau_1}^{\infty}
n ( \tau_2 - \tau_1 )  d \tau_2 \rightarrow \infty \;\; , \\
\ee

where $\tau_1$ is the present time (age of the universe) and $\tau_2 $ is some future time which is allowed to go to infinity. This is the problem of divergence raised by Hawking, \cite{hawking}. \\

\vspace{0.2in}

\noindent
{\bf Solution for a non-divergent advanced mass}\\

\noindent
 Since the universe was previously dominated by radiation it has a particle horizon $H_p$ and it is currently dominated by dark energy, because of the accelerating expansion, which implies it has an event horizon $H_e$ \cite{rindler, ajp, sciam, mnras}.  The particle horizon is the distance beyond which an observer cannot see at the current time. The event horizon is the distance beyond which the observer will never see. Using the parameter $t_0$ as the current age of the universe, the integrals maybe written as,

\bea
H_p &=&  R(t_0) \int_0^{t_0} \frac{ c dt}{R(t)} \nn \\
H_e &=&  R(t_0) \int_{t_0}^{\infty} \frac{ c dt}{R(t) } \;\; .
\eea

\noindent
Both these integrals are clearly defined Rindler \cite{rindler} and later by Ellis and Rothman \cite{ajp}. The event horizon arises because of the accelerating rate of the expansion of the universe. Rindler \cite{rindler} describes a race track with a photon like runner who travels at a constant pace, but the finish line (at the far reaches of the universe) accelerates away at a faster speed. The photon-like particle will never reach the distant universe. There is a finite distance it can cover as the time tends to infinity. This only holds for an acceleration in the expansion of the universe, which applies to our current situation...\\

\noindent
Unfortunately, the Einstein-de Sitter model (used by Hawking) does not allow for an accelerating expansion so there is no event horizon cutoff. Instead, consider a form of scale factor which allows for acceleration, for example
\be
R(t) = \left( \frac{t}{T} \right)^{3/2} \;\; .
\ee
This corresponds to a deceleration factor of $q=-1/3$, which happens to agree with current observational data, see Kumar \cite{kumar}.
This type of scale factor is treated in the text by Peebles \cite{peebles} and work on ``Power Law Inflation" by
Lucchin and Matarrese \cite{lucc}.
This gives a power law form $R(t) \propto t^n$ where $n = 1/(1+q) = 3/2$ \cite{kumar}.
The proper time in comoving coordinates becomes,
\bea
\tau &=& \int_0^t \frac{dt}{R(t)} \rightarrow 2 T^{3/2} t^{-1/2} \nn \\
t^{1/2} &=& \frac{2 T^{3/2} }{\tau} \;\;\; \mbox{ which gives } \; t^{3/2} = \frac{ 8 T^{9/2}}{\tau^3} \nn \\
R(\tau) &=& \frac{ 8 T^3}{\tau^3}
\eea
The event horizon for the scale factor $R(t)= ( t/T)^{3/2}$ is given by
\bea
H_e &=& R(\tau_1) \int_{\tau_1}^{\infty} \frac{  c dt}{R(t)} \nn \\
&=& T^{3/2} \left( \frac{\tau_1}{T} \right)^{3/2} \int_{\tau_1}^{\infty} c t^{-3/2} dt \nn \\
&=& 2 c \tau_1
\eea
hence $H_e/c = 2 \tau_1$ is the upper limit for the $m_{\mbox{\tiny adv }}$ integral.
First we evaluate $m_{\mbox{\tiny ret }}$ to find the correct form for the variable $T$ then we evaluate the
$m_{\mbox{\tiny adv }}$ integral using the same form for $T$.\\

\be
m_{\mbox{\tiny ret }}(\tau_1) = \frac{1}{R(\tau_1)} \int \frac{ N \delta ( r + \tau_2 - \tau_1 )
\Omega^3(\tau_2 ) }{ 4 \pi r } 4 \pi r^2 dr d\tau_2
\ee
where the $R$ terms from $N = nR^{-3}$ and $\Omega^3$ cancel, and the $r$ integral can be performed  using the delta function to give,
\bea
 m_{\mbox{\tiny ret }}(\tau_1) &=& -\frac{\tau_1^3 n}{8 T^3} \int_0^{\tau_1}  n (\tau_2 - \tau_1) d\tau_2 \nn \\
 &=& \frac{1}{2} n \frac{\tau_1^5}{8 T^3} = \frac{n}{L^{-3}} m
\eea
where $\tau_1 $ is the current age of the universe and the integral is over the past light cone. We have used
\be
T^3 = \left( \frac{\tau_1^5 L^{-3} }{16 m} \right)
\ee

\noindent
The advanced mass field integral over the future light cone can be written with limits from $H_e/c= 2 \tau_1$ to $\tau_1$.
\bea
m_{\mbox{\tiny adv}} (\tau_1) &=& \frac{1}{R(\tau_1)} \int \frac{ N \delta( r -\tau_2 + \tau_1) \Omega^3(\tau_2 ) }{ 4 \pi r } 4 \pi r^2 dr d\tau_2 \nn \\
&=& \frac{ n \tau_1^3}{8 T^3} \int_{\tau_1}^{2 \tau_1} ( \tau_2 - \tau_1 ) n d \tau_2 \nn \\
&=&  \frac{ n \tau_1^3}{8 T^3} \left[ \frac{\tau_2^2}{2} - \tau_2 \tau_1 \right]_{\tau_1}^{2\tau_1} \nn \\
&=& \frac{ n \tau_1^3}{8 T^3} \left( \frac{\tau_1^2}{2} \right) =  \frac{n}{L^{-3}}m \label{madv}
\eea
where we have used the same $T$ as before for the last step.
This gives $m = (1/2)( m_{\mbox{\tiny adv}} +m_{\mbox{\tiny ret}})$  since the number density of particles in the universe is  $n=L^{-3}$. Hence, we find that the advanced mass component is not infinite. \\

\noindent
In order for the scale factor $t^{3/2}$ to be realized, the universe must be undergoing a form of acceleration \cite{peebles}.
This requires that the cosmological constant be kept in the RW line element, and that the pressure in $T_{\mu\nu}$ be negative. The negative pressure comes from a  scalar field $\phi$ sometimes called an inflaton, more recently this has been compared with the Higg's field. It would be interesting to compare the scalar field in HN theory with the corresponding inflaton scalar field and see what form the potential energy of this vacuum field should take.
 The Lagrangian density satisfies the equation,
 \be
 L = \frac{1}{2} \phi_{,\mu} \phi_{,\nu} g^{\mu\nu} - V(\phi) \;\; .
 \ee
 A possible potential has been derived by Peebles \cite{peebles}, Lucchin and Matarrese \cite{lucc}. The potential in this case, $V(\phi) = A \exp( b \phi )$ where $A$ and $b$ are constants.

\subsection*{Conclusions}

\noindent
Work by Hogarth \cite{hogarth} and Hoyle and Narlikar (HN) \cite{HN64, HN, HNbook} have paved the way to a
new version of gravitational theory which is fully Machian, incorporates advanced waves fully and has Einstein's theory
as a special case. We have shown that by inclusion of the event horizon \cite{rindler} as a natural cutoff, that the advanced waves no longer yield a divergent integral. The event horizon is a manifestation of the accelerating universe and was only recently discovered \cite{riess}. Einstein understood Mach's principle, as a gravitational interaction between a test particle and the rest of the mass-energy of the universe, to be of a radiative nature and to act instantaneously as inertial forces are experienced instantly on the application of an external force. This is made possible if the Machian gravitational interaction is carried by advanced waves as in the Hoyle and Narlikar theory. HN theory reduced to Einstein's general relativity in the limit of a smooth fluid approximation and a transformation into the rest frame of the fluid.\\

\noindent
The advanced wave concept has been used successfully by Cramer in his Transactional Interpretation of quantum mechanics \cite{cramer, ruth}. HN theory has in its field equation mass fluctuation terms of the type hypothesized by Woodward \cite{jim, jimbk}. These mass fluctuations were pointed out by Fearn et al \cite{FHN}, presented at the Joint Propulsion Conference (JPC) in Ohio 2014.\\

\subsection*{Acknowledgements}

I thank J. F. Woodward for useful historical discussions for the introduction to this paper. Also Woodward pointed out to me that an accelerating universe would have an event horizon and that would most likely solve the divergence problem raised by Hawking. He was correct, it does.

\end{document}